\documentclass[%
 reprint, amsmath,amssymb,
 aps,superscriptaddress,
]{revtex4-2}

\usepackage{bm}% bold math
\usepackage{soul,xcolor}
\sethlcolor{yellow} % Change to your desired color

\usepackage{graphicx}% Include figure files
\usepackage{dcolumn}% Align table columns on decimal point
\usepackage{float}
\usepackage{footnote}
\usepackage{amsmath}
\usepackage[utf8]{inputenc}
\usepackage{array}
\usepackage{makecell}
\usepackage{gensymb}
\usepackage{adjustbox}
\usepackage{ulem}
\usepackage{color}
\usepackage[colorlinks=true, allcolors=blue]{hyperref}

\begin{document}

\preprint{APS/123-QED}

\title{Cusplike feature in Hall resistivity of a uniaxial ferromagnet in nonorthogonal Hall geometry}

\author{Banik Rai}\email{banikrai87@gmail.com}
\author{Nitesh Kumar}\email{nitesh.kumar@bose.res.in}

\affiliation {Department of Condensed Matter and Materials Physics\\S. N. Bose National Centre for Basic Sciences, Salt Lake City, Kolkata-700106, India}

\begin{abstract}
Recent magnetotransport studies on uniaxial ferromagnets have reported a cusplike feature in Hall resistivity when the magnetic field is tilted away from the conventional orthogonal direction of the Hall measurement. This feature has often been attributed to the topological Hall effect arising from a non-coplanar spin structure. In this article, we have studied the uniaxial ferromagnet SmMn$_2$Ge$_2$ to demonstrate that this feature is rather a consequence of the nonorthogonal geometry of the Hall measurement and is expected to appear whenever the magnetic field is applied away from the easy axis of magnetization, nonorthogonal to the sample plane. The Hall resistivity, exhibiting this feature, scales with the orthogonal component of the magnetization, indicating that the observed feature is simply a manifestation of the anomalous Hall effect. We explain the origin of this feature based on the evolution of ferromagnetic domains under a nonorthogonal external magnetic field.
\end{abstract}

\maketitle
\maketitle

%\tableofcontents
\section{Introduction}
When a current-carrying slab is subjected to an orthogonal magnetic field, electrons experience a Lorentz force that deflects them in a direction perpendicular to both the current and the magnetic field. This phenomenon, known as the Hall effect, was discovered by Edwin Hall in 1879\cite{Hall_effect} and has since become an indispensable tool in condensed matter physics. In ferromagnets, the electrons undergo an additional transverse deflection due to the magnetization of the sample, giving rise to the so called anomalous Hall effect (AHE).\cite{Nagaosa2010} The Hall effect is commonly expressed in terms of transverse resistivity (or Hall resistivity), $\rho_{\textit{yx}}$ (\textit{x} represents the current direction, \textit{y} represents the Hall voltage direction), as follows:
\begin{align}
    \rho_{yx} & =\rho_{yx}^\mathrm{O}+\rho_{yx}^\mathrm{A} \nonumber \\
    & =R_0B_{z}+\mu_0R_\mathrm{S}M_{z}
\label{eq1}
\end{align}
$\rho_{yx}^\mathrm{O}$ and $\rho_{yx}^\mathrm{A}$ are the ordinary and anomalous Hall resistivity arising from the Lorentz force driven ordinary Hall effect (OHE) and from the AHE, respectively. $R_0$ and $R_\mathrm{S}$ denote the ordinary and anomalous Hall coefficients, respectively. $B_{z}$ and $M_{z}$ represent the \textit{z}-components of the magnetic field and magnetization, respectively. In a conventional Hall effect experiment, the magnetic field is always applied perpendicular to the sample plane (typically the \textit{x-y} plane), making the subscript “\textit{z}” in $B_z$ redundant. However, for an arbitrary direction of the magnetic field, the subscript ``\textit{z}" becomes significant as only the out-of-plane (orthogonal) component ($B_z$) of the magnetic field contributes to the OHE. Although the in plane component of the magnetic field does not contribute to the OHE, it can however give rise to the even-in-field planar Hall effect (PHE),\cite{PHE} which originates from the anisotropic magnetoresistance of the sample and is typically very small. The PHE vanishes when the in-plane component of magnetic field aligns with the current direction. An in-plane field generally induces an in-plane magnetization, ensuring that $M_z = 0$, thereby preventing the AHE as well. Nonetheless, a finite AHE can, in principle, arise even with an in-plane field, though this is rare due to stringent symmetry requirements.\cite{EuCd2Sb2,ZrTe5,Fe3Sn}

From Eq. (\ref{eq1}), it is evident that the shape of the field-dependent curve of $\rho_{yx}$ closely resembles the field-dependent curve of $M_z$, as OHE is typically small compared to AHE and appears only as a weak linear background in the field-dependent curve of $\rho_{yx}$. This resemblance is often broken in materials hosting non-coplanar spin structures, such as skrmions, which impart an additional transverse velocity to electrons through real-space Berry curvature mechanisms.\cite{THE} This phenomenon, known as the topological Hall effect (THE), serves as a key signature of non-coplanar spin textures in a material. 

Recent studies on uniaxial ferromagnets, including Fe$_3$GaTe$_2$,\cite{Fe3GaTe2} Fe$_3$GeTe$_2$,\cite{Fe3GeTe2_PRB,Fe3GeTe2_Scientific_reports}  Cr$_{1.2}$Te$_2$,\cite{Cr1.2Te2} \textit{R}Mn$_2$Ge$_2$ (\textit{R}= La,\cite{LaMn2Ge2} Ce,\cite{CeMn2Ge2} Pr,\cite{PrMn2Ge2} Sm\cite{SmMn2Ge2}), have reported a cusplike feature in the field-dependent Hall resistivity when the magnetic field is tilted away from the conventional orthogonal direction. This feature, apparently absent in the corresponding magnetization data, has largely been attributed to the THE arising from the non-coplanar spin structure. However, we identify a fundamental shortcoming in the interpretation of the magnetization data in these studies. When the magnetic field deviates from the conventional orthogonal direction, the magnetization also becomes nonorthogonal to the sample plane. Conventional magnetization measurements, often performed using vibrating sample magnetometers (VSM) or superconducting quantum interference devices (SQUID), measure the component of magnetization parallel to the applied magnetic field ($M_{B||}$). However, the AHE is driven by the orthogonal component of magnetization ($M_{z}$). In the cited studies, the reported magnetization corresponds to $M_{B||}$ and not to $M_{z}$, leading to the apparent discrepancy between the Hall resistivity and magnetization data.

In this study, we employ VSM and torque magnetometry techniques to determine $M_{z}$ under a nonorthogonal magnetic field in SmMn$_2$Ge$_2$, a uniaxial ferromagnet. We show that the cusplike feature is present in the field dependence of $M_{z}$ also. This observation indicates that the feature in Hall resistivity is merely a manifestation of the AHE driven by $M_{z}$ in nonorthogonal Hall geometry rather than a THE signature. Furthermore, we provide a plausible origin of this feature based on the evolution of ferromagnetic domains under a nonorthogonal external magnetic field.

\begin{figure}[t]
\includegraphics[width=\linewidth]{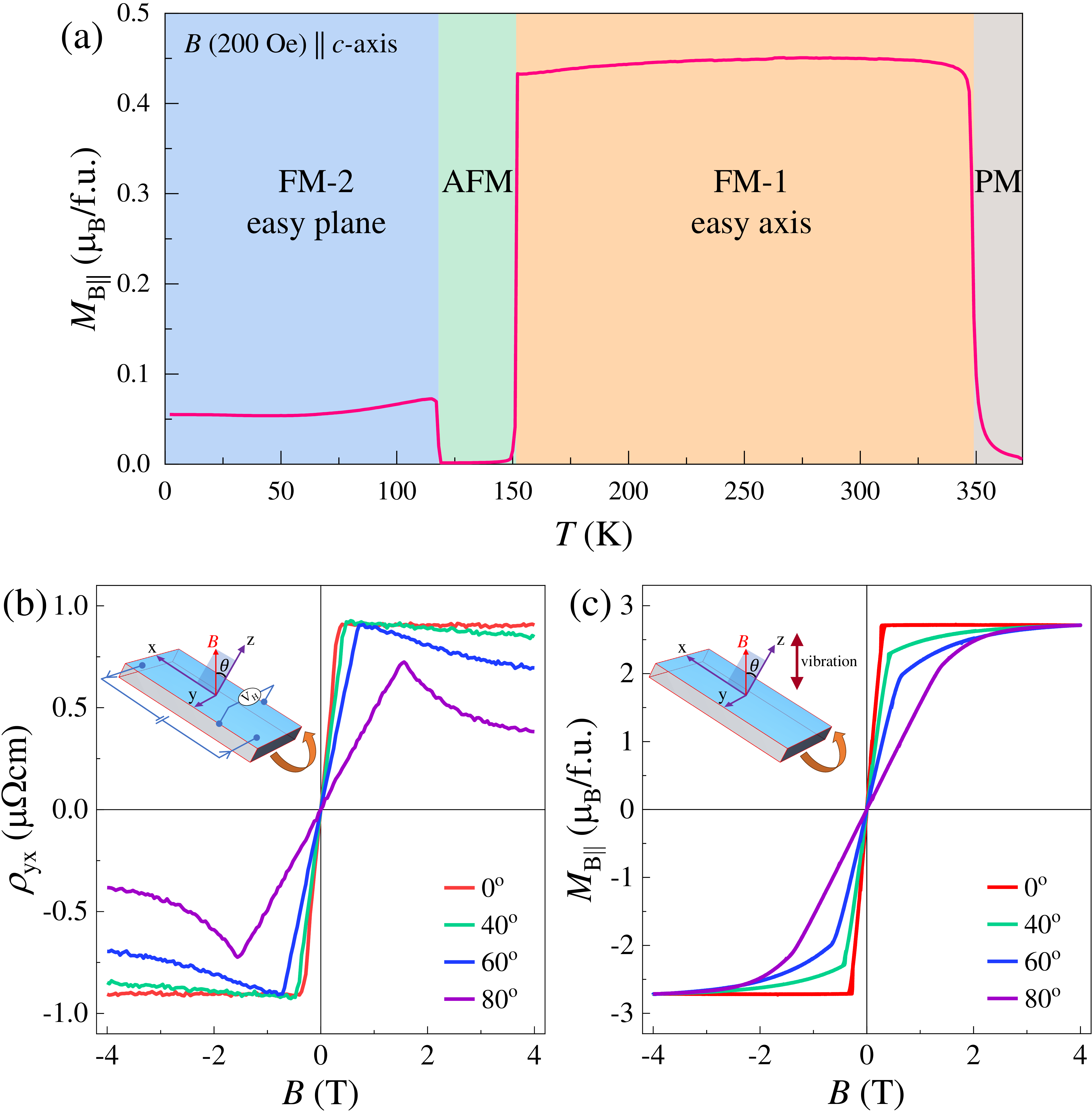}
\caption{\label{fig1} (a) Temperature-dependent magnetization with the magnetic field (200 Oe) applied along the \textit{c}-axis. Different magnetic phases are shown. (b)-(c) Magntic field dependent (b) Hall resistivity and (c) magnetization for various magnetic field tilts relative to the \textit{z}-direction}
\end{figure}

\begin{figure}[t]
\includegraphics[width=\linewidth]{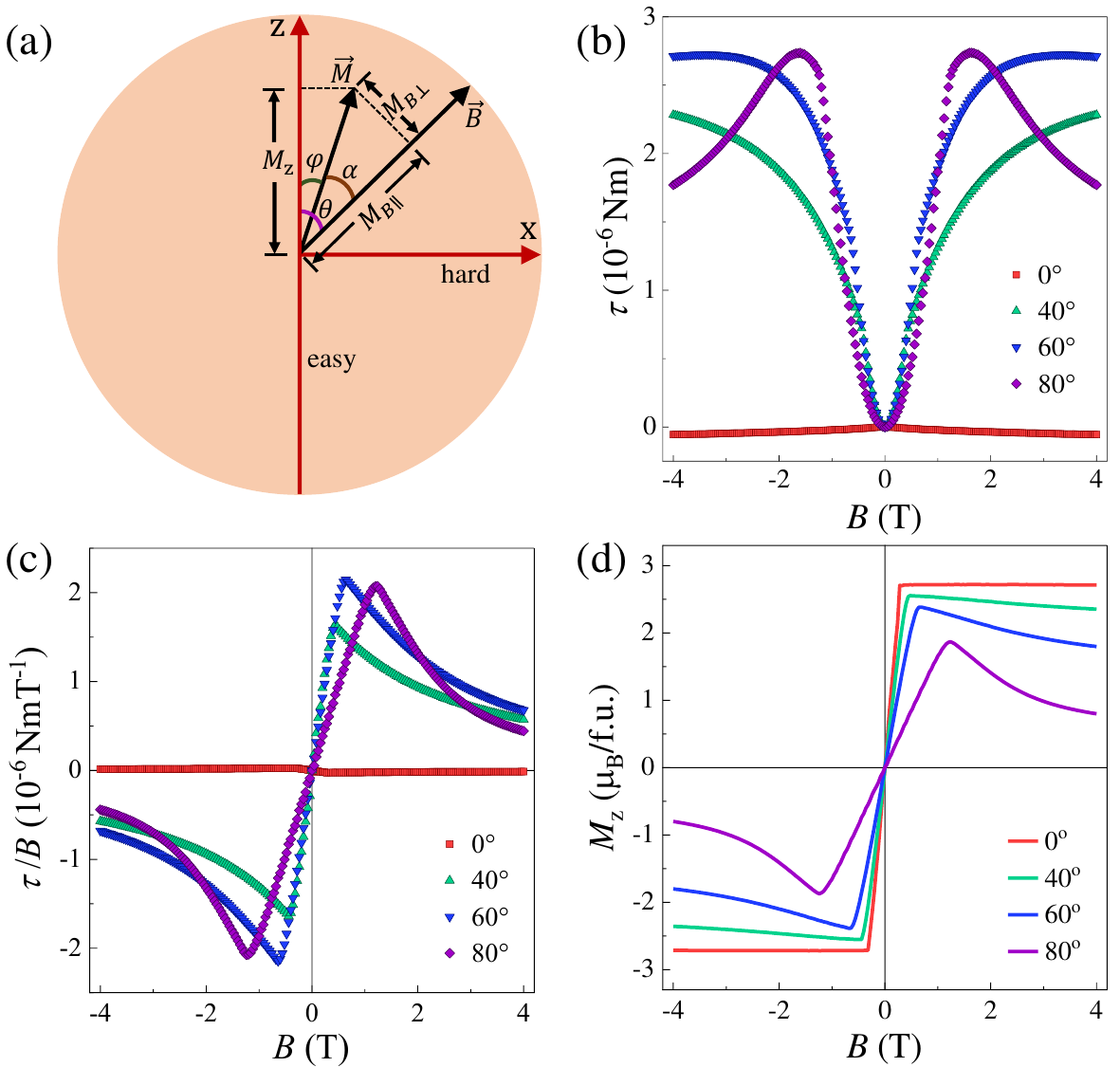}
\caption{\label{fig2} (a) A schematic showing the various components of magnetization vector for the tilt $\theta$ of the magnetic field relative to the \textit{z}-axis. The easy and hard axes of magnetization are also shown. (b) Torque ($\tau$) acting on the sample as a function of magnetic field for various values of $\theta$. (c) Variation of the quantity $\tau/B$ as a function of magnetic field for various values of $\theta$. (d) Variation of the z-component of the magnetization ($M_z$) as a function of the magnetic field for various values of $\theta$.}
\end{figure}
\section{Results and discussion}
SmMn$_2$Ge$_2$ crystallizes in a tetragonal structure (space group $I4/mmm$) and undergoes multiple magnetic transitions as the temperature is lowered, as shown in Fig. \ref{fig1}(a).\cite{Fujii1985} Above 350 K, it is paramagnetic. Between 350 K and 150 K, the compound is ferromagnetic (FM-1), with magnetic moments spontaneously aligned along [001] (\textit{c}-axis). From 150 K to 120 K, it transitions to an antiferromagnetic state. Below 120 K, SmMn$_2$Ge$_2$ reverts to a ferromagnetic state again (FM-2), with the magnetic moments aligned in the basal-plane along [110]. For our purpose, we choose the FM-1 state, where SmMn$_2$Ge$_2$ is a uniaxial ferromagnet, with the \textit{c}-axis being the easy axis of magnetization. Fig. \ref{fig1}(b) and \ref{fig1}(c) present the field-dependent Hall resistivity and magnetization, respectively, for various tilt angles ($\theta$) of the magnetic field relative to the \textit{z}-axis. Here, the \textit{z}-axis corresponds to the \textit{c}-axis, while the \textit{x}- and \textit{y}-axes lie along the \textit{a}-axis of the tetragonal unit cell of SmMn$_2$Ge$_2$. For $\theta = 0^\circ$ (no tilt), the Hall resistivity initially increases steeply with the applied magnetic field and then saturates, a typical behavior in orthogonal Hall geometry. At finite tilt angles, the Hall resistivity initially increases linearly with the magnetic field, reaches a maximum, and then decreases gradually, forming a cusplike feature. This feature becomes more pronounced as the tilt angle increases. Notably, the field-dependent magnetization measured under the same tilt conditions does not exhibit any cusplike features. This discrepancy arises because the data presented in Fig. \ref{fig1}(c) is not the orthogonal component ($M_z$) of the magnetization but the component that is parallel to the applied magnetic field ($M_{B||}$). Fig. \ref{fig2}(a) shows the schematic diagram of different components of magnetization for the given tilt of the magnetic field. When the magnetic field is not sufficiently strong to fully polarize the sample, the magnetization vector lies somewhere between the \textit{z}-axis and the direction of the magnetic field, making angles $\phi$ and $\alpha$ with the former and latter, respectively, such that $\phi +\alpha=\theta$. The \textit{z}-component of the magnetization is given by $M\cos\phi$, which can be expressed in terms of $M_{B||}$, $M_{B\perp}$ (the component perpendicular to the magnetic field) and $\theta$ as follows:
\begin{align}  
    M_z&=M\cos{\phi}\nonumber\\
    &=M\cos(\theta-\alpha)\nonumber\\
    &=M(\cos\theta\cos\alpha+\sin\theta\sin\alpha)\nonumber\\
    &=M_{B||}\cos\theta+M_{B\perp}\sin\theta
\label{eq2}    
\end{align}

\begin{figure}[t]
\includegraphics[width=\linewidth]{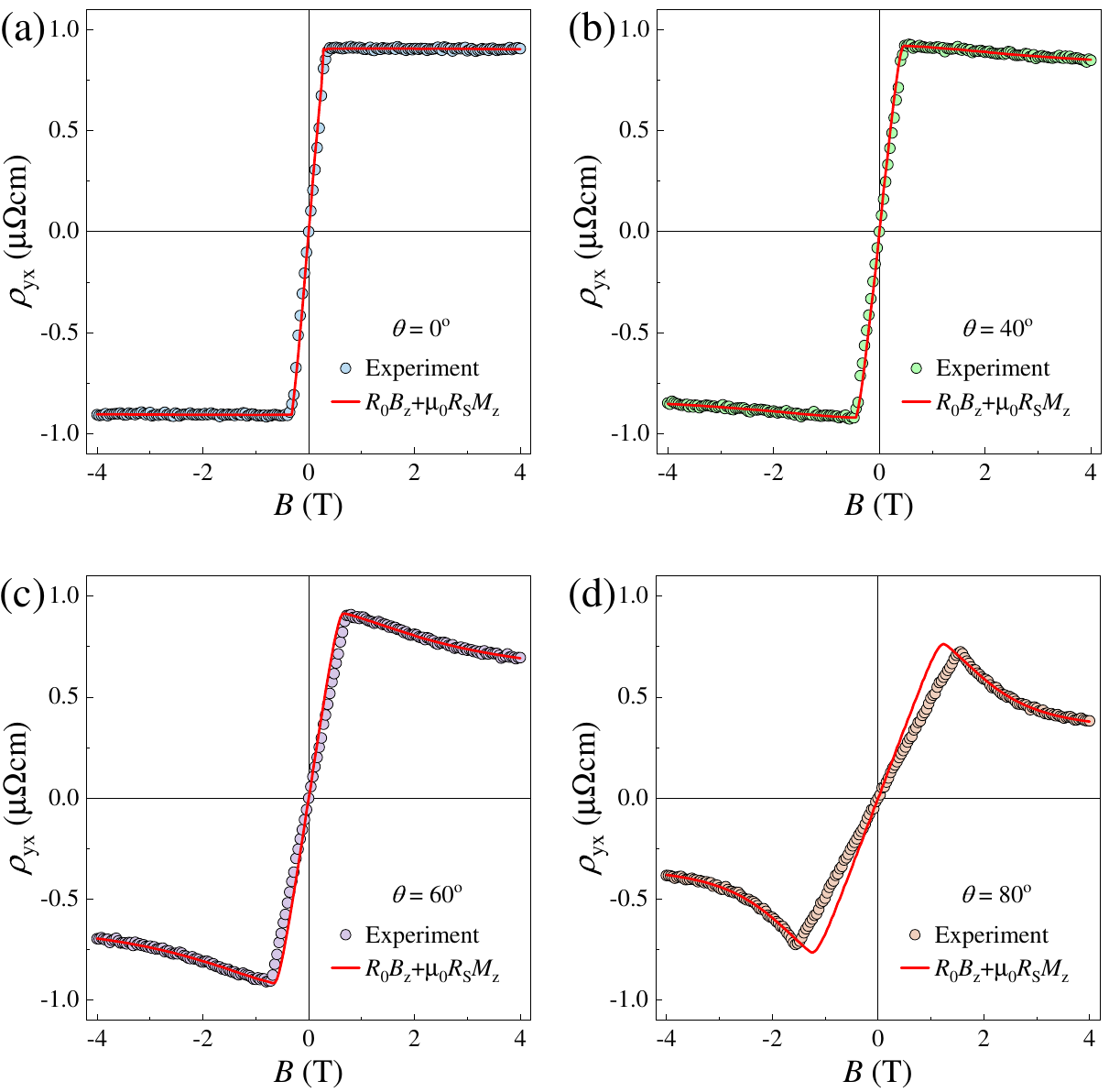}
\caption{\label{fig3} Fitting of the Hall resistivity for various tilts ($\theta$) of the magnetic field relative to the z-axis, considering the combined contributions of ordinary and anomalous Hall resistivity, as given by Eq. (\ref{eq1}).}
\end{figure}

The only unknown term in Eq. (\ref{eq2}) is $M_{B\perp}$. To evaluate this, we utilize the torque magnetometry. The torque acting on the sample for the given configuration is given by $\tau=M_{B\perp}B$. By measuring the torque as a function of the magnetic field, the component $M_{B\perp}$ can be directly determined using $M_{B\perp}=\tau/B$. Figure \ref{fig2}(b) shows the torque acting on the sample as a function of the magnetic field for various values of $\theta$. For $\theta = 0^\circ$, no torque is expected since the magnetization and magnetic field are aligned in the same direction. However, a small negative torque is observed, likely due to the slight misalignment of the sample during measurement. Fig. \ref{fig2}(c) shows the field dependence of the quantity $\tau/B$ for various values of $\theta$. Fig. \ref{fig2}(d) shows the field dependence of $M_z$, calculated using Eq. (\ref{eq2}), for various values of $\theta$. The $M_z$ shown in Fig. \ref{fig2}(d) exhibits cusplike features similar to those observed in the Hall resistivity in Fig. \ref{fig1}(a). This strongly suggests that the cusplike feature in Hall resistivity is the manifestation of AHE, driven by $M_z$ rather than a THE signal.

Fig. \ref{fig3} presents the Hall resistivity fitted with Eq. (\ref{eq1}), with the value of $M_z$ calculated from Eq. (\ref{eq2}). The fitting shows good agreement for lower values of $\theta$. At higher values of $\theta$, the quality of the fitting deteriorates, which could be attributed to two potential factors. A slight misalignment of the sample during measurement can alter the value of $\theta$. A more probable cause is that, as the torque increases with the magnetic field, the cantilever of the torque magnetometry chip may rotate, effectively reducing the value of $\theta$. This results in the \textquotedblleft cusp\textquotedblright appearing at lower magnetic fields, as seen in Fig. \ref{fig3}(d). 

\begin{figure}[b]
\includegraphics[width=\linewidth]{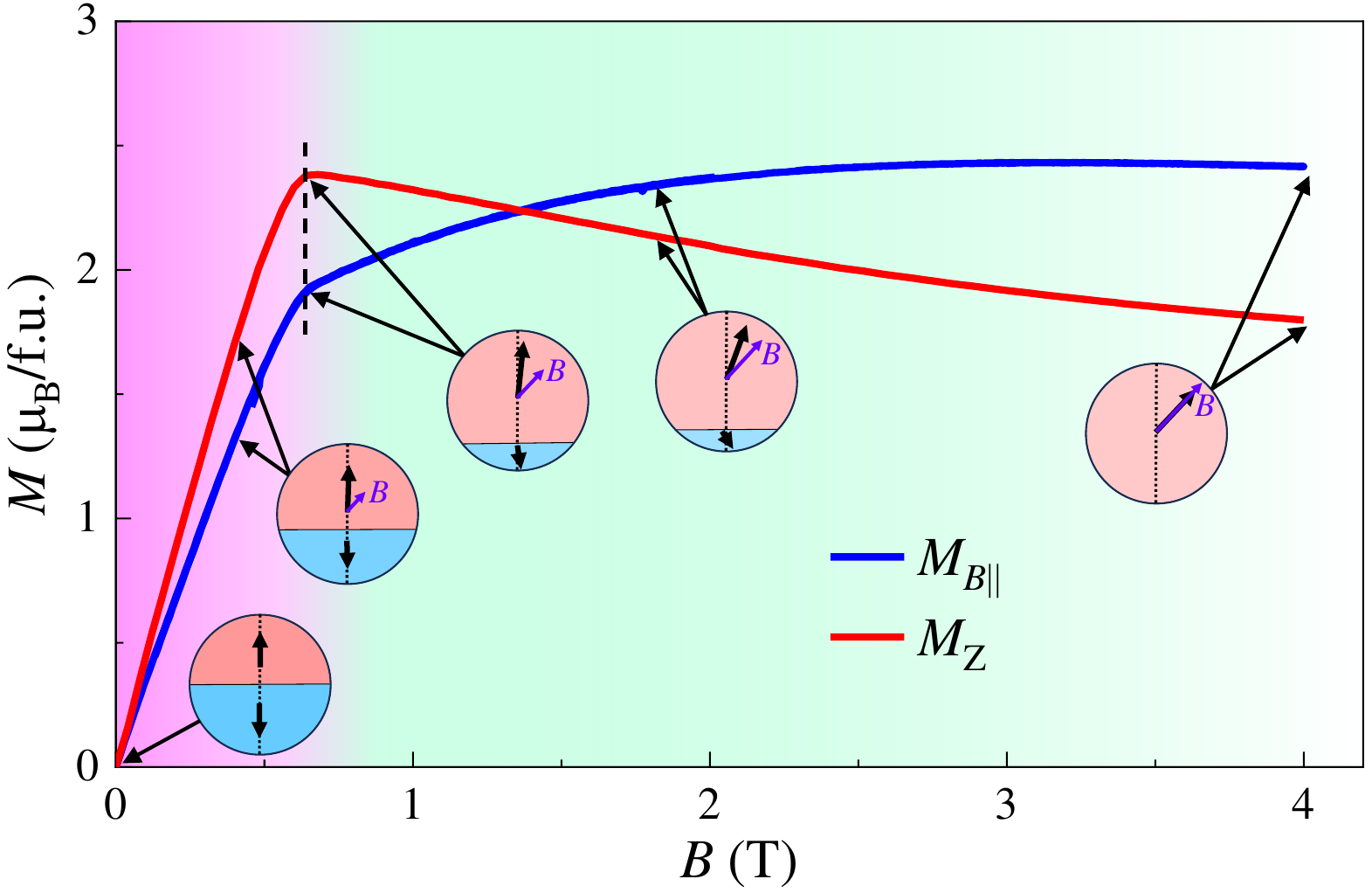}
\caption{\label{fig4} {Variation of the magnetization components $M_z$ and $M_{B||}$ with the magnetic field applied at an angle $\theta=60^\circ$ to the $z$-axis. A possible domain evolution is shown for various magnetic field strengths.}}
\label{fig4}
\end{figure}

Next, we aim to explain the cusplike feature in $M_z$ considering the evolution of ferromagnetic domains under a nonorthogonal magnetic field. Due to magnetocrystalline anisotropy, the domains in uniaxial ferromagnets tend to align along the easy axis of magnetization. In the absence of magnetic field, the net magnetization is zero, which is achieved by the formation of two oppositely aligned domains, parallel to the easy axis, whose net moments cancel each other (see Fig. \ref{fig4}). When a magnetic field is applied at an angle to the easy axis (the $z$-axis in our case), the domain inclined toward the field grows in size, while the domain inclined away from the field shrinks, driven by domain-wall motion. At lower fields, domain rotation remains negligible due to the strong influence of magnetocrystalline anisotropy, and the magnetization is primarily governed by domain-wall motion. Nonetheless, both $M_z$ and $M_{B||}$ increase sharply as the magnetic field is increased from zero, as shown in Fig. \ref{fig4}. As the field continues to increase, domain wall motion approaches completion, and domain rotation becomes dominant. As the domain rotates toward the field, i.e., away from the $z$-axis, the component $M_{B||}$ continues to increase with the field, while $M_z$ starts to decrease. This trend persists until the magnetic field becomes large enough to complete the domain rotation, at which point the sample becomes fully polarized, and both $M_z$ and $M_{B||}$ reach saturation. This results in a cusplike feature in the field dependence of $M_z$, which is subsequently reflected in the Hall resistivity measured with the same tilt of the magnetic field. Since this analysis is applicable to all uniaxial ferromagnets whenever the magnetic field is tilted away from the easy axis of magnetization, a similar trend can be expected in other uniaxial ferromagnets as well.

\section{Conclusion}
We studied the prototype uniaxial ferromagnet SmMn$_2$Ge$_2$ to understand the origin of the cusplike feature observed in the field-dependent Hall resisitivity under a nonorthogonal magnetic field. We showed that this feature, commonly observed in many uniaxial ferromagnets, arises as a simple consequence of the nonorthogonal Hall geometry and is a mere manifestation of the anomalous Hall effect, driven by the orthogonal component of the magnetization. The cusplike feature emerges as long as the out-of-plane (orthogonal) component of the magnetic field is non-zero. Therefore,  even in seemingly in-plane field configurations ($\theta = 90^\circ$), a slight misalignment of the sample, often encountered in experiments, can introduce a nonzero out-of-plane component of the magnetic field, leading to the cusplike feature in the Hall resistivity as observed in Refs \cite{Fe3GaTe2,Cr1.2Te2,SmMn2Ge2}. Given the increasing number of reports erroneously attributing this feature to the topological Hall effect, this study aims to clarify its actual origin and prevent future misinterpretations.

\section{Method}The single crystals of SmMn$_2$Ge$_2$ were grown by the flux method, the details of which are given in Ref. \cite{singh_prm}. All measurements, including electrical transport, magnetization, and torque magnetometry, were performed at 200 K in a Physical Property Measurement System (PPMS, Quantum Design, Dynacool, 9T). The sample was rotated using the standard rotator probe of the PPMS for transport and torque measurements, whereas custom-made quartz wedges, precisely cut at the required angles, were used to achieve the tilted magnetic field configurations in magnetization measurements. The Hall resistivity was measured using a standard four-probe method. Torque measurements were conducted using the torsional cantilever technique, while magnetization was measured using the vibrating sample magnetometry (VSM) technique, which detects the magnetization component parallel to the applied magnetic field.

\textit{Note added}. A qualitative discussion on the cusplike feature in Hall resistivity of SmMn$_2$Ge$_2$ has been given by us in Ref. \cite{singh_prm}.

\section*{Acknowledgements}This research project utilized the instrumentation facilities of the Technical Research Centre (TRC) at the S. N. Bose National Centre for Basic Sciences, under the Department of Science and Technology (DST), Government of India. NK acknowledges the Science and Engineering Research Board (SERB), India, for financial support through Grant Sanction No. CRG/2021/002747 and Max Planck Society for funding under Max Planck-India partner group project. BR acknowledges financial support from the DST through the senior research fellowship.

\section*{Data availability}The data are available from the authors on a reasonable request.

%\bibliography{references}% Produces the bibliography via BibTeX.
\bibliographystyle{unsrt} 
\end{document}